\documentclass[twocolumn]{autart}

\usepackage{graphicx,graphics}
\usepackage{amssymb,amsmath}  
\usepackage{dsfont}
\usepackage{enumerate}
\usepackage{cite}

\usepackage{savesym}
\savesymbol{AND}
\savesymbol{OR}
\savesymbol{NOT}
\savesymbol{TO}
\savesymbol{COMMENT}
\savesymbol{BODY}
\savesymbol{IF}
\savesymbol{ELSE}
\savesymbol{ELSIF}
\savesymbol{FOR}
\savesymbol{WHILE}

\usepackage{algorithm,algorithmic}

\newtheorem{theorem}{Theorem}

\newtheorem{lemma}[theorem]{Lemma}

\begin{document}
\begin{frontmatter}

\title{
Guaranteed sensor coverage with the \emph{weighted-}$D^2$ sampling
}
\author[IBM]{Ajay Deshpande}\ead{ajayd@us.ibm.com},
\address[IBM]{IBM T. J. Watson Research Center, Yorktown Heights, NY, USA}  

\begin{abstract}
In this paper we focus on the mobile sensor coverage problem formulated as a continuous locational optimization problem. Cort\`{e}s et al. first proposed a distributed version of the Lloyd descent algorithm with guaranteed convergence to a local optima. Since then researchers have studied a number of variations of the coverage problem. The quality of the final solution with the Lloyd descent depends on the initial sensor configuration. Inspired by the recent results on a related $k$-means problem, in this paper we propose the weighted-$D^2$ sampling to choose the initial sensor configuration and show that it yields $O(\log k)$-competitive sensor coverage before even applying the Lloyd descent. Through extensive numerical simulations, we show that the initial coverage with the weighted-$D^2$ sampling is significantly lower than that with the uniform random initial sensor configuration. We also show that the average distance traveled by the sensors to reach the final configuration through the Lloyd descent is also significantly lower than that with the uniform random configuration. This also implies considerable savings in the energy spent by the sensors during motion and faster convergence.
\end{abstract}

\end{frontmatter}

\section{Introduction}
\label{sec:intro}
\emph{Sensor coverage} is a fundamental issue in large-scale sensing. It addresses the question of when and where to place the sensors. With the advent of miniaturized wireless sensors, sensing is now possible at unprecedented scales. Naturally, sensor coverage has received significant attention in the last decade and to date many different formulations have been proposed. In all the formulations, sensor coverage formalizes a metric for the \emph{quality-of-sensing} and the sensor coverage problem aims to find an optimal or a near-optimal sensor placement which optimizes the metric.

In this paper we focus on a sensor coverage problem that is formulated as a \emph{locational optimization problem} as in \cite{cortes2002coverage,cortes2004coverage}. Locational optimization problems have long been studied in various fields in the context of static spatial resource allocation and solved in a centralized computing environment \cite{okabe1997locational,cortes2004coverage}. Cort\`{e}s et al. for the first time formulated the mobile sensor coverage problem as a locational optimization problem and developed a framework for distributed control and coordination of mobile sensors to obtain optimal coverage \cite{cortes2002coverage,cortes2004coverage}. A \emph{centroidal Voronoi configuration}, where each sensor is at the centroid of its Voronoi partition, is well-known to be an optimal solution to this problem \cite{okabe1997locational}. Cort\`{e}s et al. proposed a distributed \emph{Lloyd gradient descent} algorithm with the guaranteed convergence to a centroidal Voronoi configuration. According to this control algorithm, starting with an initial configuration each sensor incrementally moves towards the centroid of its Voronoi partition until the convergence \cite{cortes2002coverage,cortes2004coverage}. This work initiated the study of several variations of the mobile sensor coverage problem, for example, coverage with limited range constraints for sensing and communication \cite{cortes2005spatially}, coverage with a generalized monotonic sensing function \cite{schwager2007distributed}, coverage with sensor consensus and learning \cite{schwager2009decentralized}, and coverage with sensors having location-dependent sensing performance \cite{deshpande2009distributed}. Every work focused on developing a distributed control law for sensors movement, typically as an extension of the Lloyd descent algorithm, and proved the guaranteed convergence to a locally optimal solution.

An important consideration that is seldom addressed in this area is the quality of the (locally) optimal solution obtained with the Lloyd descent. It depends on the initial sensor configuration. In the works so far, the initial sensor configuration is typically chosen uniformly at random. There is no guarantee on the optimal solution thus obtained. In this paper, we propose \emph{weighted-$D^2$} sampling to choose the initial sensor configuration, which guarantees \emph{$O(\log k)$-competitive} sensor coverage before even applying the Lloyd descent algorithm. Here, $k$ denotes the number of sensors. The application of the Lloyd descent only improves the sensor coverage. Our work is inspired based on the recent work \cite{Arthur:2007:KAC:1283383.1283494} by Arthur and Vassilvitskii on using $D^2$ sampling for choosing the initial centers in the related \emph{discrete $k$-means problem} for clustering with the $O(\log k)$-competitive guarantee. We prove our result in two steps. In the first step, we establish a close relationship between the sensor coverage problem and a suitably selected discrete weighted $k$-means problem. In the second step, we extend the original result in \cite{Arthur:2007:KAC:1283383.1283494} to the weighted-$D^2$ sampling for the weighted $k$-means problem. To the best of our knowledge, this is a first foray into the performance guarantees for sensor coverage before applying any control algorithm. In \cite{salapaka2003constraints}, the authors proposed a deterministic annealing approach adapted to the locational optimization problem to obtain a globally optimal solution. However, there is no guarantee on the convergence rate.

The weighted-$D^2$ sampling is computationally straightforward. We first create a discrete set of candidate sensor locations using a grid and assign a weight to each location. Then we choose the first sensor location at random from the discrete candidate set with the probability proportional to the weight. Subsequently, we iteratively select the next sensor location with the probability proportional to the weight times the square of the shortest distance to the sensor locations already chosen. Through extensive numerical simulations, we observe that with the weighted-$D^2$ sampling, the initial sensor coverage before applying the Lloyd descent is significantly lower than that with the uniform random deployment.

Another advantage of the weighted-$D^2$ sampling for the initial configuration is the energy spent by mobile sensors moving during the Lloyd descent.  While recent works (e.g. \cite{kwok2007energy, derenick2011energy}) aim to develop energy balancing Lloyd descent control laws, this applies once the initial configuration is chosen. The weighted-$D^2$ sampling already with the guarantee on the initial sensor coverage helps in significantly lowering the energy spent during the descent. Through our simulations, we find that with the weighted-$D^2$ sampling, the average distance traveled by the sensors to the convergence during the Lloyd descent is considerably lower than that with the uniform random initial configuration. This also means that the energy spent is also significantly lower and depending on the sensor motion dynamics, the convergence is faster.

\textbf{Organization:} In Section \ref{sec:related work} we discuss the related work to the sensor coverage and the $k$-means problems. In Section \ref{sec:coverage control problem and solution} we define our sensor coverage problem, discuss the Lloyd descent and propose the weighted-$D^2$ sampling method for choosing the initial sensor configuration. In Section \ref{sec:weighted k-means and coverage problem}, we establish the relationship between the sensor coverage and the discrete weighted $k$-means problem. In Section \ref{sec:gurantees on weighted D2} we prove the guarantees on the solution to the weighted $k$-means problem with the weighted-$D^2$ sampling. In Section \ref{sec: numerical results}, we show present results of the numerical simulations. In Section \ref{sec: conclusion} we conclude and outline future work.


\section{Related work}
\label{sec:related work}
Locational optimization problems have long been studied in the areas of spatial economics and facility location \cite{okabe1997locational,cortes2004coverage}. Okabe and Suzuki present a review of a class of continuous locational optimization problems which can be addressed using Voronoi diagrams \cite{okabe1997locational}. Cort\`{e}s et al. formulated the mobile sensor coverage problem as a continuous locational optimization problem for the first time \cite{cortes2002coverage,cortes2004coverage}. We will discuss their formulation in detail in the next section. They proposed a

Since then researchers have studied several variations of the mobile sensor coverage. Examples include coverage using sensors with limited-range sensing and communication \cite{cortes2005spatially},  coverage with a generalized monotonic sensing function \cite{schwager2007distributed}, coverage with sensor consensus and learning \cite{schwager2009decentralized}, and coverage over networks using sensors with location-dependent sensing performance \cite{deshpande2009distributed}. All these works have developed variations of the Lloyd descent algorithm and proved guaranteed convergence to local optimal in a distributed setting. In recent years, there has also been some work on developing control laws that balance coverage as well as energy across sensors \cite{kwok2007energy, derenick2011energy}. In all these works, the initial sensor placement is chosen uniformly at random. In \cite{deshpande2013optimal} Deshpande et al. developed a version of the Lloyd descent algorithm for the sensor placement in physical networks such as water distribution networks. Using simulations, they showed that the initial sensor configuration with the $D^2$ sampling yields an optimal solution with better coverage than the uniform random placement, converges faster, and is also significantly closer to the globally optimal solution. In all these works there is no guarantee on the quality of the optimal solution. In \cite{salapaka2003constraints}, the authors proposed a deterministic annealing approach adapted to a class of locational optimization problems to obtain globally optimal solutions. However, there is no guarantee on the convergence rate.

A related problem in the discrete world is the $k$-means problem which aims to choose $k$ centers for a given set of points such that the sum of the squared distance between each point and its closest center is minimized. This naturally partitions the points into $k$ clusters with the points in the same cluster being the closest to the same center. The center for each cluster turns out to be the centroid of the points in that cluster. $k$-means is widely used as a clustering technique in unsupervised learning. While $k$-means problem is known to be \emph{NP-hard}, a locally optimal solution can be obtained very fast using the Lloyd descent algorithm. In \cite{Arthur:2007:KAC:1283383.1283494}, Arthur and Vassilvitskii introduce $D^2$-sampling to choose $k$ initial points and show that the solution is already $O(\log k)$-competitive. We will discuss $D^2$-sampling in detail in this paper.


\section{Coverage Problem and weighted-$D^2$ sampling}
\label{sec:coverage control problem and solution}
In this section we first introduce the sensor coverage problem and discuss the Lloyd descent algorithm. Then we propose the weighted-$D^2$ sampling.

\subsection{Coverage problem formulation}
\label{subsec:coverage problem formulation}
We consider the same coverage problem formulation in 2D as in \cite{cortes2004coverage}.
Let $Q$ denote a bounded convex region in $\mathbb{R}^2$. Let $\phi(\cdot)$ denote a nonnegative scalar density function $\phi:Q\rightarrow\mathbb{R}_{+}$. $\phi(q)$ can be viewed as a measure of importance of covering location $q$. Without loss of generality we assume that $\int\limits_{Q} \phi(q)dq = 1$. Let $P = (p_1,p_2,\cdots,p_k)$ denote the locations of $k$ mobile sensors in $Q$. The sensing performance at location $q$ due to the $i$th sensor located at $p_i$ degrades with the Euclidean distance $\|q-p_i\|$ between $q$ and $p_i$. We assume that it is modeled as $\|q-p_i\|^2$. For fixed sensor locations $(p_1,p_2,\cdots,p_k)$, the sensing performance function induces a \emph{Voronoi partition} $\mathcal{V}(P) = \{V_1,V_2,\cdots,V_k\}$ of $Q$, where,
\begin{equation}\label{eqn:voronoi parition}
    V_i = \{q ~|~ \|q-p_i\|^2 \le \|q-p_j\|^2 ~~\forall j\ne i\}.
\end{equation}
The sensor located at $p_i$ \emph{covers} all points in $V_i$. For given sensor locations $P$, the coverage cost or metric $H(P)$ is given by:
\begin{equation}\label{eqn:coverage metric}
    H(P) = \sum_{i=1}^k \int\displaylimits_{V_i} \|q-p_i\|^2\phi(q)dq.
\end{equation}
The optimal sensor coverage problem is to find an optimal sensor configuration $P^\ast = (p_1^\ast,p_2^\ast,\cdots,p_k^\ast)$ such that the coverage metric is minimized.
\begin{equation}\label{eqn:optimal coverage}
    P^\ast = \arg\min_{P} H(P).
\end{equation}
This problem is also referred to as a \emph{continuous k-median problem} \cite{cortes2004coverage}, \cite{Fekete:2000:CWK:336154.336182}.

\subsection{Lloyd descent for coverage control}
\label{subsec:lloyd algorithm}
In \cite{cortes2004coverage} Cort\'{e}s et al. show that \emph{centroidal Voronoi configuration}, where each sensor is at the centroid of its own Voronoi partition, is a (locally) optimal solution to the above coverage problem. Further, they
propose continuous and discrete-time variations of Lloyd's descent algorithm to control movements of mobile sensors to reach a centroidal Voronoi configuration starting from an initial configuration. In these algorithms, a sensor moves towards the center of its Voronoi partition. As the sensors move, Voronoi partitions as well as their centroids evolve. Cort\'{e}s et al. prove that the sensors converge to an optimal configuration. During the sensor location update, the algorithm ensures that the coverage metric reduces. The authors further propose an asynchronous distributed control scheme and prove its convergence as well. The quality of the optimal solution depends on the initial sensor configuration. There are no guarantees on the locally optimal coverage cost with respect to the global optimum.

\subsection{Weighted-$D^2$ sampling for initial sensor configuration}
\label{subsec:weighted D2 sampling on discrete space}
\begin{figure}
\centering
  \includegraphics[width = 3.4in]{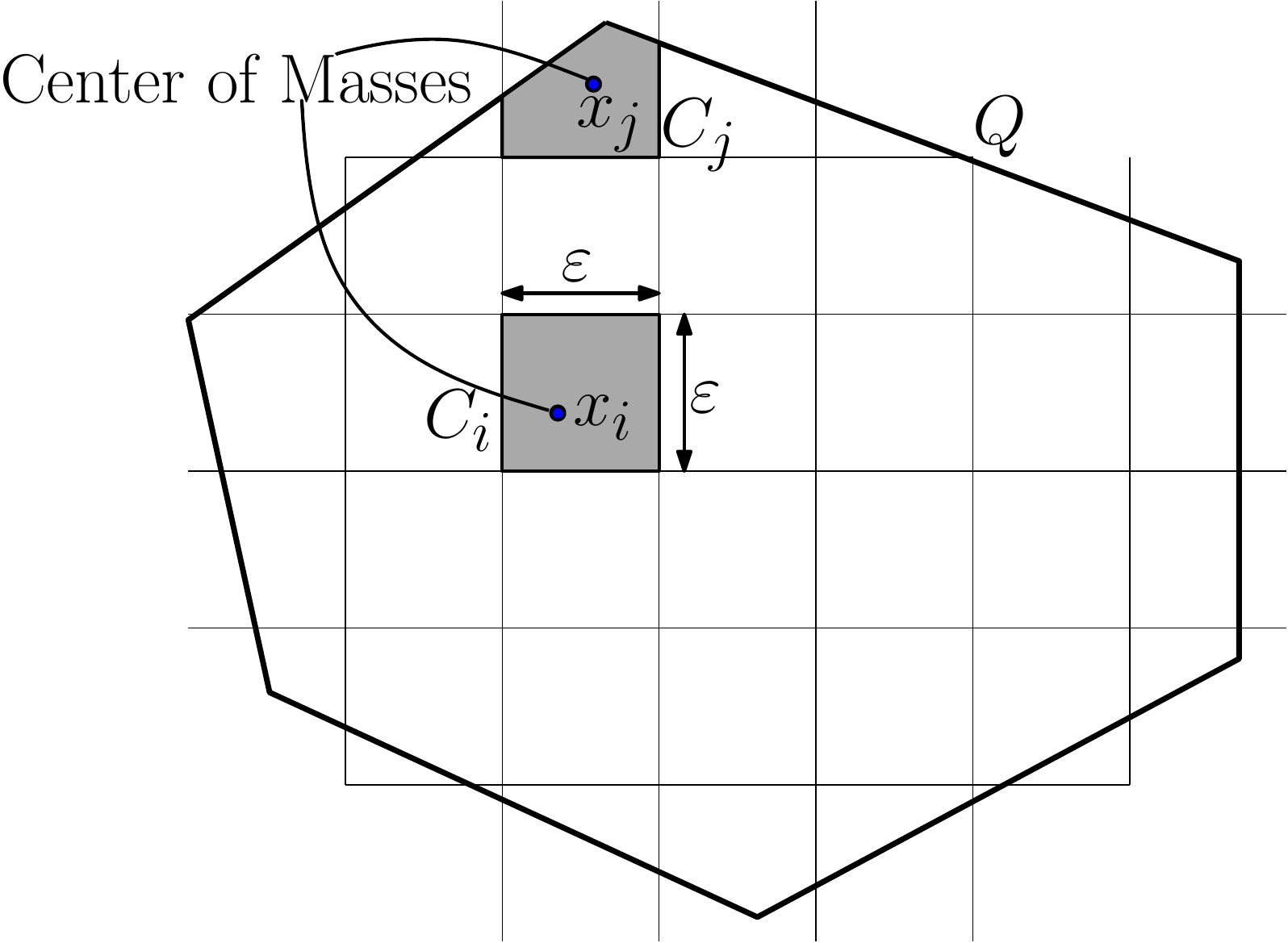}
  \caption{Superimposition of uniform square grid of size $\varepsilon \times \varepsilon$ on the convex domain $Q$ to generate a cell partition. Cell lying entirely within $Q$ are square grid cells. Cells on the boundary of $Q$ form convex polygonal cells.}
  \label{fig:grid impose}
\end{figure}

We propose a weighted-$D^2$ sampling approach to select initial sensor locations. It consists of two steps. In the first step, we discretize the domain $Q$ using a uniform square grid and select the \emph{centers} of cells as candidate locations. Next we use a special sampling procedure to select $k$ initial sensor locations. As shown in Figure \ref{fig:grid impose}, we superimpose a uniform square grid of size $\varepsilon \times \varepsilon$ on $Q$ and create $C_1, C_2, \cdots, C_n$ cells. Note that some cells are entire square grid cells, whereas the others on the boundary of $Q$ form convex polygons. We define the \emph{weight} $w_i$ of each cell $C_i$ as follows:
\begin{equation}\label{eqn:cell weight}
    w_i = \int\displaylimits_{C_i} \phi(q)dq.
\end{equation}
Since we assume that $\int\limits_{Q} \phi(q)dq = 1$,
\begin{equation}\label{eqn:normalize density}
    \sum\limits_{i=1}^n w_i = 1.
\end{equation}

We compute the \emph{center of mass} $x_i$ of each cell $C_i$ as follows:
\begin{equation}\label{eqn:cell center of mass}
    x_i = \frac{1}{w_i}\int\displaylimits_{C_i} q\phi(q).
\end{equation}
We consider $X = \{x_1, x_2, \cdots, x_n\}$ as a set of candidates for the choice of $k$ initial sensor locations. Next we use an iterative sampling procedure to select $k$ locations from $X$. Let $D(x)$ denote the shortest distance of point $x$ to the sensor locations already chosen.
\begin{enumerate}
  \item Select the first sensor location $p_1$ as $x_i$ at random from $X$ with the probability proportional to $w_i$.
  \item Select the new location as $x_i$ at random from $X$ with the probability $\frac{w_iD(x_i)^2}{\sum_i w_iD(x_i)^2}$. Update $D(x_i)$.
  \item Repeat Step (2) until we have chosen $k$ locations.
\end{enumerate}

We refer to the above procedure as weighted-$D^2$ sampling. It is similar to the $D^2$-sampling proposed in \cite{Arthur:2007:KAC:1283383.1283494} except that there is an additional weighing with $w_i$'s. Once we select $k$ initial sensor locations, we continue to apply the Lloyd descent algorithm. We prove the following theorem which is the central result in this paper.

\begin{theorem}
\label{thm:main theorem}
Let $P = (p_1,p_2,\cdots,p_k)$ denote the set of $k$ initial sensor locations obtained by applying the above weighted-$D^2$ sampling procedure. Then,
\begin{equation}
\mathrm{E}[H(P)] \le 8(\ln k + 2)H(P) + 16\sqrt{2}(\ln k + 2)D\varepsilon + \sum_{i} J_{C_i|x_i},
\end{equation}
where $J_{C_i|x_i}$ is the moment of inertia of cell $C_i$ about its center of mass $x_i$, and $D$ is the maximum distance between two Voronoi neighbors.
\end{theorem}

For a sufficiently granular grid division $\varepsilon$ the last two terms in the above statement can be ignored and bounded by a small constant. This result implies that the coverage obtained with the weighted-$D^2$ sampling is $O(\log k)$-competitive. We prove the above theorem in three parts. In the first part, we establish a relationship between the coverage metric $H(P)$ and the related coverage metric for a weighted $k$-means problem. In the second part, we show that the weighted-$D^2$ sampling provides performance guarantees on the coverage metric for the weighted $k$-means problem. In the third part, we combine the results of the first two parts to prove the main result.

\section{Weighted $k$-means problem and its relation to the coverage problem}
\label{sec:weighted k-means and coverage problem}

We first define the weighted $k$-means problem in generality. Let $X = \{x_1,x_2,\cdots,x_n\}$ be a set of $n$ points in $\mathbb{R}^d$. Each $x_i$ has an associated non-negative scalar weight $w_i \ge 0$. The weighted $k$-means problem involves choosing a set of $k$ points $P = (p_1, p_2, \cdots, p_k)$ from $\mathbb{R}^d$ such that $\Phi(P) = \sum\limits_{x_i\in X} \min\limits_{p \in P} w_i\|x_i-p\|^2$ is minimized. The difference between the weighted $k$-means problem and the $k$-means problem (\emph{e.g.} \cite{Arthur:2007:KAC:1283383.1283494}) is the weighting by $w_i$'s.

Now we consider the weighted $k$-means problem in the context of the coverage problem setting that we discussed in the previous section. Let $X = \{x_1,x_2,\cdots,x_n\}$ be the set of the center of masses (Equation \ref{eqn:cell center of mass}) of cells $C_i$'s as we constructed by the grid discretization of $Q$ and let $w_i$'s be the weights as defined by Equation \ref{eqn:cell weight}. Now we establish the relationship between the metric $\Phi(P)$ for the weighted $k$-means problem and the coverage metric $H(P)$ given by the following theorem.

\begin{theorem}
\label{thm:k-means and coverage}
\begin{equation}\label{eqn:relation between k-means and coverage}
    H(P) \le \Phi(P) + \sum_{i} J_{C_i|x_i} \le H(P) + 2\sqrt{2}D\varepsilon,
\end{equation}
where $J_{C_i|x_i}$ is the moment of inertia of cell $C_i$ about its center of mass $x_i$, and $D$ is the maximum distance between two Voronoi neighbors.
\end{theorem}

\emph{Proof}:
Let $h(C_i)$ and $\phi(C_i)$ denote the contributions associated with cell $C_i$ to the metrics $H(P)$ and $\Phi(P)$ respectively. Thus, $H(P) = \sum\limits_i h(C_i)$ and $\Phi(P) = \sum\limits_i \phi(C_i)$, where,

\begin{equation}\label{eqn: cell coverage metric}
    h(C_i) = \int\displaylimits_{q\in C_i} \phi(q)\min_{p\in P}(\|q-p\|^2)dq,
\end{equation}
and,
\begin{equation}\label{eqn: weighted k-means metric}
    \phi(C_i) = w_i\min_{p\in P}(\|x_i-p\|^2).
\end{equation}

\begin{figure}
\centering
  \includegraphics[width = 3.4in]{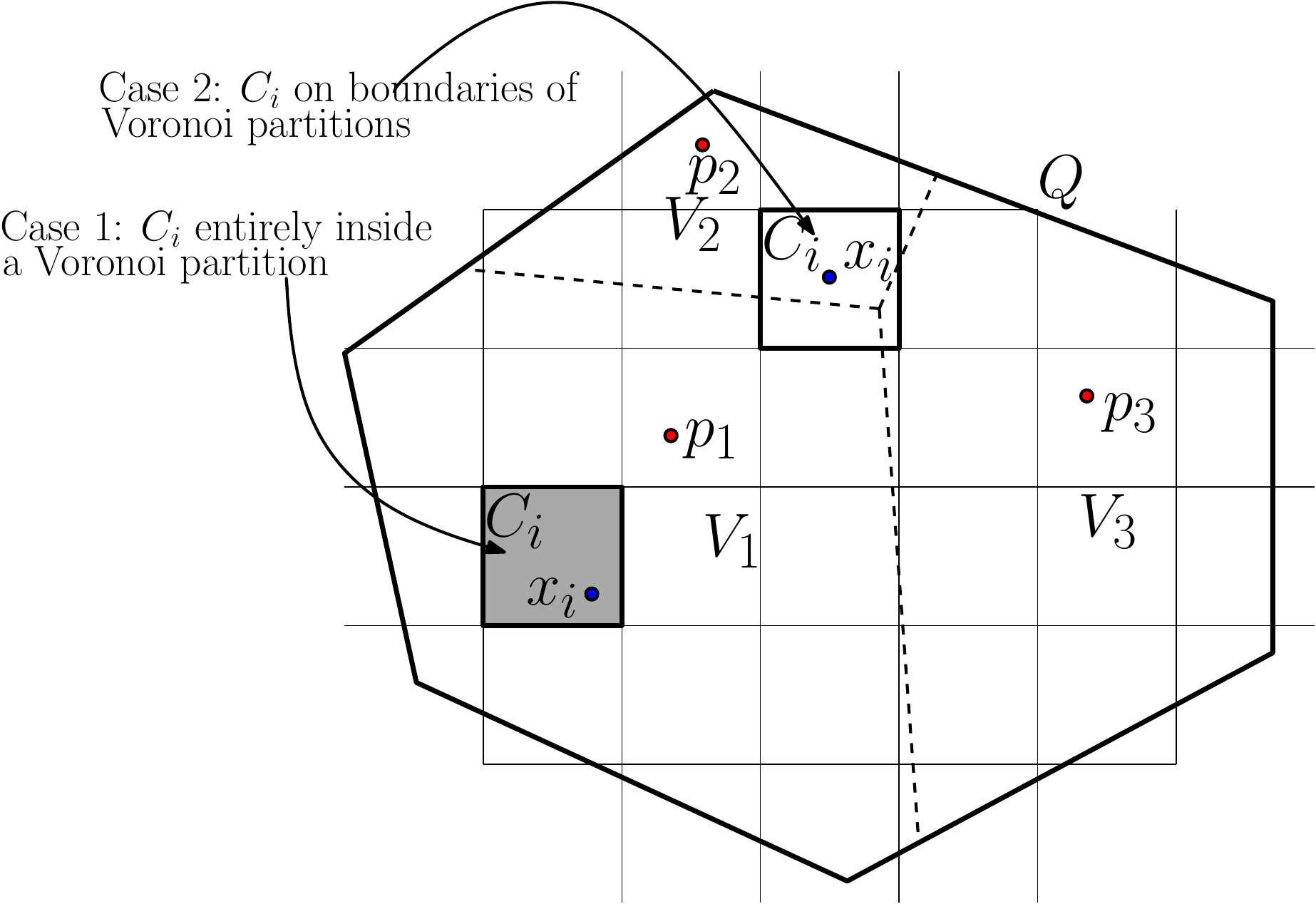}
  \caption{Two cases based on whether (1) a cell lies entirely inside a Voronoi partition of some sensor location or (2) a cell lies on the boundaries of two or more Voronoi partitions}
  \label{fig:voronoi}
\end{figure}

\begin{figure}
\centering
  \includegraphics[width = 3in]{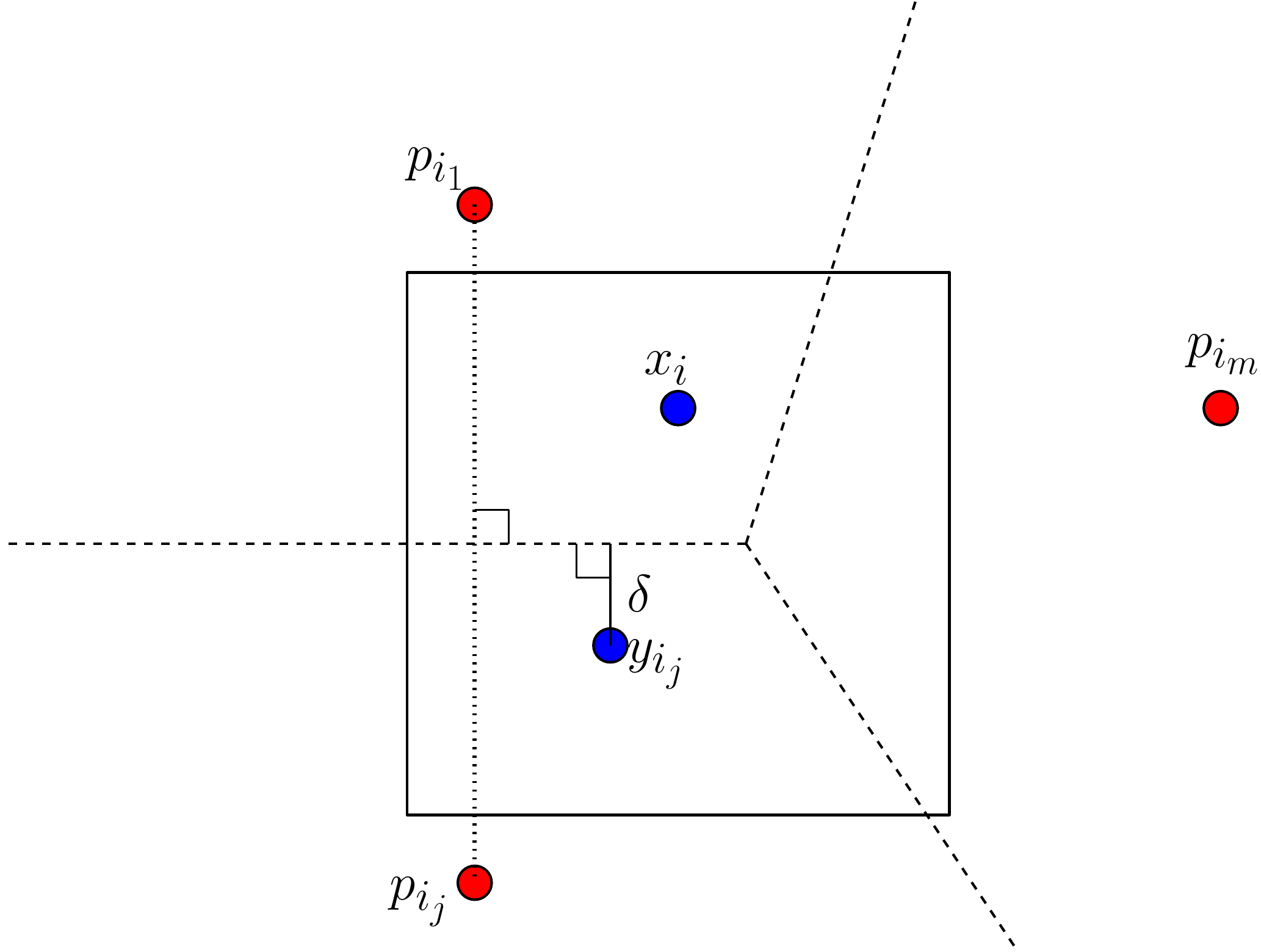}
  \caption{A detailed illustration of a cell getting divided by two or more Voronoi partitions.}
  \label{fig:boundary cell}
\end{figure}

Now we establish the relationship between $h(C_i)$ and $\phi(C_i)$. As discussed in Equation \ref{eqn:voronoi parition}, let $V_1, V_2, \cdots, V_k$ denote the Voronoi partition associated with $p_1, p_2, \cdots, p_k$. Now we consider two cases as shown in Figure \ref{fig:boundary cell}. In the first case, a cell lies entirely within a Voronoi partition for some sensor location. In the second case, a cell lies on the boundaries of two or more Voronoi partitions.

\emph{Case 1}: Suppose $C_i$ lies entirely within some Voronoi partition. Let $V_j$ denote that partition for point $p_j$. Then, $\phi(C_i) = w_i\|x_i-p_j\|^2$ and,
\begin{equation}\label{eqn: cell coverage metric}
    h(C_i) = \int\displaylimits_{q\in C_i} \phi(q)\|q-p_j\|^2dq.
\end{equation}
By parallel axis theorem,
\begin{equation}\label{eqn: cell coverage metric M.I.}
    h(C_i) = w_i\|x_i-p_j\|^2 + J_{C_i|x_i},
\end{equation}
where $J_{C_i|x_i} = \int\limits_{q\in C_i} \phi(q)\|q-x_i\|^2dq$. $J_{C_i|x_i}$ is essentially the moment of inertia of cell $C_i$ about its center of mass $x_i$. Thus,
\begin{equation}\label{eqn: inner cell coverage relation}
    h(C_i) = \phi(C_i) + J_{C_i|x_i}.
\end{equation}

\emph{Case 2}: $C_i$ lies on the boundaries of two or more Voronoi partitions. Suppose $C_i$ lies at the boundaries of Voronoi partitions $V_{i_1}, V_{i_2}, \cdots, V_{i_m}$ corresponding to the points $p_{i_1}, p_{i_2}, \cdots, p_{i_m}$. Without loss of generality, suppose $x_i$ lies in $V_{i_1}$. Let $D_{i_j} = C_i \cap V_{i_j}$ denote the part of $C_i$ belonging to $V_{i_j}$. Further, let $y_{i_j}$ denote the center of mass of each $D_{i_j}$, given by
\begin{equation}\label{eqn:piece center of mass}
    y_{i_j} = \frac{1}{v_{i_j}}\int\displaylimits_{D_{i_j}}q\phi(q)dq,
\end{equation}
where weight $v_{i_j} = \int\displaylimits_{D_{i_j}}\phi(q)dq$. Note that $w_i = \sum_{j=1}^{m}v_{i_j}$.
Now, $\phi(C_i) = w_i\|x_i-p_{i_1}\|^2$.

\begin{eqnarray}
\nonumber  h(C_i) &=& \sum_{j=1}^{m}\int\displaylimits_{q\in D_{i_j}} \phi(q)\|q-p_{i_j}\|^2dq \\
\nonumber   &=& \sum_{j=1}^{m}\left[v_{i_j}\|y_{i_j}-p_{i_j}\|^2 + J_{D_{i_j}|y_{i_j}}\right] \\
\nonumber   &=& \sum_{j=1}^{m}\left[v_{i_j}\|y_{i_j}-p_{i_j}\|^2 - v_{i_j}\|y_{i_j}-p_{i_1}\|^2\right] \\
\nonumber   &+& \sum_{j=1}^{m}\left[v_{i_j}\|y_{i_j}-p_{i_1}\|^2 + J_{D_{i_j}|y_{i_j}}\right] \\
\nonumber   &=& \sum_{j=1}^{m}v_{i_j}\left[\|y_{i_j}-p_{i_j}\|^2 - \|y_{i_j}-p_{i_1}\|^2\right] \\
\nonumber   &+& w_i\|x_i-p_{i_1}\|^2 + J_{C_i|x_i}\\
\nonumber   &=& \sum_{j=1}^{m}v_{i_j}\left[\|y_{i_j}-p_{i_j}\|^2 - \|y_{i_j}-p_{i_1}\|^2\right] + \phi(C_i) + J_{C_i|x_i}.
\end{eqnarray}

Rearranging the above,

\begin{equation}\label{eqn:boundary cell coverage relation}
    h(C_i) + \sum_{j=1}^{m}v_{i_j}\left[\|y_{i_j}-p_{i_1}\|^2 - \|y_{i_j}-p_{i_j}\|^2\right] = \phi(C_i) + J_{C_i|x_i}.
\end{equation}

Note that each $\|y_{i_j}-p_{i_1}\| \ge \|y_{i_j}-p_{i_j}\|$. Therefore,

\begin{equation}\label{eqn:boundary cell coverage bound 1}
    h(C_i) \le \phi(C_i) + J_{C_i|x_i}.
\end{equation}

Now we bound the second term on the left hand side in Equation \ref{eqn:boundary cell coverage relation} above. As shown in Figure \ref{fig:boundary cell}, the Voronoi partition is a perpendicular bisector of segment $P_{i_1}P_{i_j}$. Let $\delta$ be the length of the perpendicular projection of $y_{i_j}$ on the Voronoi partition crossing $P_{i_1}P_{i_j}$, which is the perpendicular bisector of $P_{i_1}P_{i_j}$. Suppose $|P_{i_1}P_{i_j}| = \|p_{i_1}-p_{i_j}\| = d_{ij}$.
\begin{eqnarray*}
  \|y_{i_j}-p_{i_1}\|^2 - \|y_{i_j}-p_{i_j}\|^2 &=& \left(\frac{d_{ij}}{2}+\delta\right)^2 + \left(\frac{d_{ij}}{2}-\delta\right)^2 \\
  &=& 2d_{ij}\delta \\
\end{eqnarray*}

Note that $\delta \le \sqrt{2}\varepsilon$. Let $D$ denote the distance between the maximally separated Voronoi neighbors. Thus,
$\|p_{i_1}-p_{i_j}\| = d_{ij} \le D$. Substituting these bounds in the above equation, we obtain,

\begin{equation}
\label{eqn:distance bound}
  \|y_{i_j}-p_{i_1}\|^2 - \|y_{i_j}-p_{i_j}\|^2 = 2d\delta \le 2\sqrt{2}\varepsilon D.
\end{equation}

Substituting Equation \ref{eqn:distance bound} into Equation \ref{eqn:boundary cell coverage relation},
\begin{eqnarray}
\nonumber    \phi(C_i) + J_{C_i|x_i} &\le& h(C_i) + \sum_{j=1}^{m}v_{i_j}2\sqrt{2}\varepsilon D \\
 &\le& h(C_i) + 2\sqrt{2}w_i\varepsilon D \label{eqn:boundary cell coverage bound 2}
\end{eqnarray}

Combining Equation \ref{eqn: inner cell coverage relation} in Case 1 and Equation \ref{eqn:boundary cell coverage bound 1} in Case 2, and summing over all cells, we obtain,

\begin{equation}
\label{eqn:coverage bound 1}
    H(P) \le \Phi(P) + \sum_{i} J_{C_i|x_i}.
\end{equation}

Combining Equation \ref{eqn: inner cell coverage relation} in Case 1 and Equation \ref{eqn:boundary cell coverage bound 2} in Case 2, and summing over all cells, we obtain,

\begin{equation}
\label{eqn:coverage bound 2}
    \Phi(P) + \sum_{i} J_{C_i|x_i} \le H(P) + 2\sqrt{2}D\varepsilon\sum_{i\in \mathcal{B}(V)} w_i,
\end{equation}

where $\mathcal{B}(V)$ denotes the set of cells that are on the boundaries of the Voronoi partition. Equation \ref{eqn:normalize density} implies that $\sum_{i\in \mathcal{B}(V)} w_i \le 1$. Substituting above, we obtain:

\begin{equation}
\label{eqn:coverage bound final}
    \Phi(P) + \sum_{i} J_{C_i|x_i} \le H(P) + 2\sqrt{2}D\varepsilon.
\end{equation}

Equations \ref{eqn:coverage bound 1} and \ref{eqn:coverage bound final} imply the main result of the theorem.\qed

\section{Guarantees on the weighted $k$-means solutions with weighted-$D^2$ sampling}
\label{sec:gurantees on weighted D2}

In this section, we show that the weighted-$D^2$ sampling yields $O(\log k)$-competitive solution to the weighted $k$-means problem. Note that this result is not just applicable to the specific weighted $k$-means instance we considered in the previous section, but holds true in general. We essentially extend the result in \cite{Arthur:2007:KAC:1283383.1283494} for $D^2$ sampling. Let $\Phi_{\mathrm{OPT}}$ denote an optimal solution to $\Phi(P)$. We prove the following result.

\begin{theorem}
\label{thm:k-means guarantee}
The weighted-$D^2$ sampling procedure in Section \ref{subsec:weighted D2 sampling on discrete space} applied to the weighted $k$-means problem yields the following guarantee:
\begin{equation}
\mathrm{E}[\Phi(P)] \le 8(\ln k + 2)\Phi_{\mathrm{OPT}}.
\end{equation}
\end{theorem}

Since the proof of the above theorem exactly follows and extends the steps in \cite{Arthur:2007:KAC:1283383.1283494}, we have included the proof in Appendix \ref{sec:appendix proof weighted k-means}. Now we are ready to prove Theorem \ref{thm:main theorem}.

\emph{Proof of Theorem \ref{thm:main theorem}}:
Using the left hand side bound in Theorem \ref{thm:k-means and coverage} and Theorem \ref{thm:k-means guarantee},
\begin{eqnarray*}
\mathrm{E}[H(P)] &\le& \mathrm{E}[\Phi(P)] + \sum_{i} J_{C_i|x_i} \\
 &\le& 8(\ln k + 2)\Phi_{\mathrm{OPT}} + \sum_{i} J_{C_i|x_i}.
\end{eqnarray*}
$\Phi_{\mathrm{OPT}} \le \Phi(P^\ast)$, where $P^\ast$ is the optimal solution to $H(P)$ as in Equation \ref{eqn:optimal coverage}. Substituting above and also using the right hand side bound in Theorem \ref{thm:k-means and coverage},
\begin{eqnarray*}
\mathrm{E}[H(P)] &\le& 8(\ln k + 2)\Phi(P^\ast) + \sum_{i} J_{C_i|x_i} \\
&\le& 8(\ln k + 2)H(P) + 16\sqrt{2}(\ln k + 2)D\varepsilon + \sum_{i} J_{C_i|x_i}.
\end{eqnarray*}
\qed

\section{Numerical Results}
\label{sec: numerical results}

We performed numerical simulations to compare the sensor coverage performance of the initial sensor configuration chosen with the weighted-$D^2$ sampling to that with the uniform random initial placement. In particular, for these two types of initial configurations, we compare the initial coverage, the final coverage after applying the Lloyd descent, and the distance traveled by sensors during the descent, which is closely related to the energy spent during the descent. We performed our simulations in Matlab. We set the domain $Q$ as a unit square with the vertices at $[0,0], [1,0], [1,1]$ and $[0,1]$. We consider the density function $\phi(x,y)$ as a superposition of two Gaussians given by:
\begin{eqnarray}
\nonumber \phi(x,y) &=& \frac{1}{A}e^{-10(x-0.75)^2-2(y-0.75)^2} \\
&+& \frac{1}{A}e^{-20(x-0.25)^2-2(y-0.25)^2}
\end{eqnarray}
The centers of two Gaussians are located at $(0.25,0.25)$ and $(0.75,0.75)$. $A = 0.610882$ is the normalization constant such that $\int_Q \phi(x,y)dxdy = 1$. In order to obtain the initial sensor configuration with the weighted-$D^2$ sampling, we discretize the domain into grid cells of size $\varepsilon \times \varepsilon$ and sample $k$ locations from the set of the center of masses of the grid cells according to the algorithm in Section \ref{subsec:weighted D2 sampling on discrete space}. For the case of the initial configuration with the uniform random sampling, we choose $k$ locations uniformly at random from the unit square. We present the results for the three scenarios. In the first scenario, we set $k = 10$ and choose $\varepsilon = 0.1$, i.e., we choose a grid division of $10 \times 10$ cells. In the second scenario, we set $k = 10$, but choose $\varepsilon = 0.05$ with $20 \times 20$ grid cells to understand the effect of discretization. In the last scenario, we increase the number of sensors to $k = 20$ with $\varepsilon = 0.05$. In each scenario, we performed 50 simulation runs for each type of the initial sensor configuration. In each run, we apply the Lloyd descent until the sensors converge to the final configuration. In each iteration of the Lloyd descent, every sensor moves towards the center of its Voronoi cell with the proportional gain of $K = 10$. We choose the convergence criterion as the mean $L1$-norm of the sensor position change during each iteration being less than 1E-4. During each iteration, we compute the sensor coverage and also keep track of the distances traveled by the sensors. We discuss our observations below.

\begin{figure}
\centering
  \includegraphics[width = 3.4in]{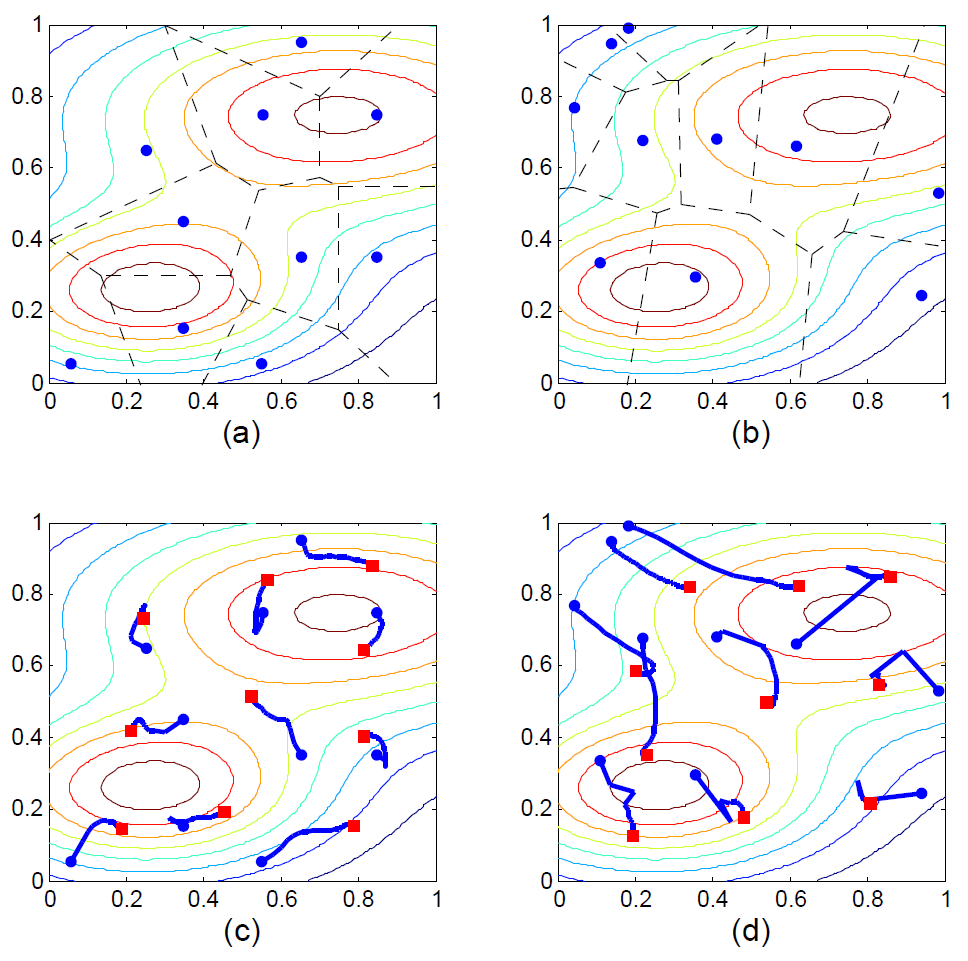}
  \caption{An illustrative example to show the comparison of the initial sensor configurations and their Voronoi partitions for the weighted-$D^2$ (Figure a) and the uniform random sampling cases (Figure b). Figures (c) and (d) also show the trajectories of the sensors during the Lloyd descent from the initial sensor configuration (blue circles) to the final configuration (red squares). Figure (c) and (d) show the weighted-$D^2$ and the uniform random sampling cases respectively.}
  \label{fig:illustrative comparison}
\end{figure}

\begin{figure}
\centering
  \includegraphics[width = 3.4in]{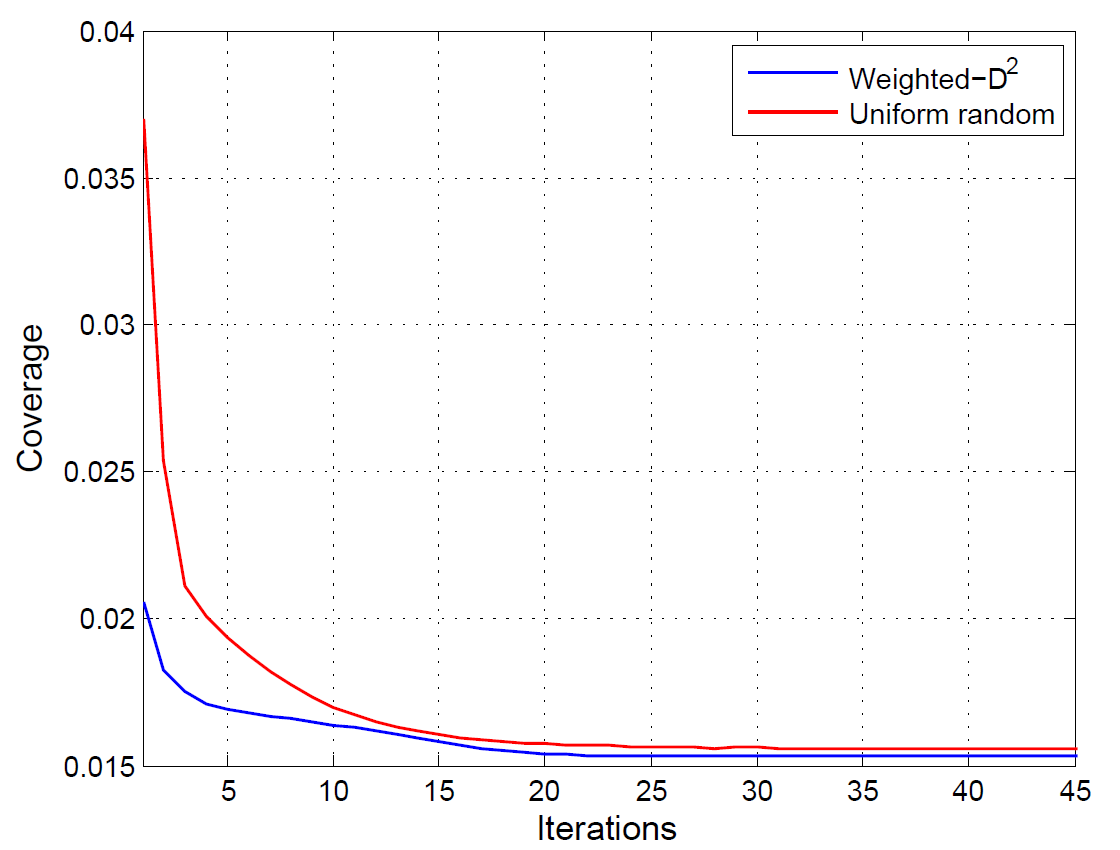}
  \caption{An illustrative example to show the comparison of the coverage value updates during the Lloyd descent for the weighted-$D^2$ and the uniform random sampling cases.}
  \label{fig:coverage during descent}
\end{figure}

As an illustration, from the runs for the first scenario of $k = 10$ and $\varepsilon = 0.1$, Figure \ref{fig:illustrative comparison} (a) and (b) respectively show an example initial sensor configuration for the weighted-$D^2$ sampling case and the uniform random case. In each case, the contours of the underlying field density function $\phi(\cdot)$ are shown. The Voronoi diagrams of the sensor locations are also shown. Figure \ref{fig:coverage during descent} shows the coverage for each case during the iterations of the Lloyd descent. In this example, the initial coverage for the weighted-$D^2$ case is much lower than the uniform random case. We consistently observe this trend across multiple runs as we will discuss shortly. In both cases, across the iterations, the final coverage is close to each other. However, for the unform random deployment, in the initial round of iterations each sensor travels much longer distance. Figure \ref{fig:illustrative comparison} (c) and (d) show the trajectories of the sensor locations in each case during the Lloyd descent. The red squares represent the final sensor configuration after the convergence. We note that the paths traveled in the uniform random case are considerably longer than the weighted-$D^2$ sampling case, and again, we consistently observe this trend across multiple runs.

In Table \ref{table:results1} and Table \ref{table:results2}, we present the summary statistics across multiple runs for each of the three scenarios considered here. Table \ref{table:results1} shows, for the cases of weighted-$D^2$ and uniform random deployments, the average initial coverage, the average final coverage, the standard deviations across multiple runs and also the \% improvement in the initial coverage with the weighted-$D^2$ sampling over the uniform random sampling. For all the scenarios, both types of deployments lead to similar final coverage values through the Lloyd descent. For each scenario, the initial coverage with the weighted-$D^2$ sampling is about 1.5 times the final coverage value, whereas for the uniform random case, the initial coverage is 2.3 times the final coverage value. In each scenario, the weighted-$D^2$ sampling yields on average over 30\% improvement in the initial coverage than the uniform random deployment. Between the first and second scenario, we note that the initial and final coverage values for the weighted-$D^2$ sampling are not too different indicating that the grid size granularity of $\varepsilon = 0.1$ is sufficient compared to $\varepsilon = 0.05$.

Table \ref{table:results2} shows for each case of the deployment the comparison of the average across multiple runs for the average distance traveled by each sensor during the Lloyd descent. The table also shows the standard deviations across the runs and the \% improvement in the average distance traveled per sensor with the weighted-$D^2$ sampling over the uniform random case. We note that the average distance traveled per sensor is reduced by at least 30\% for $k = 10$ with the weighted-$D^2$ sampling over the uniform sampling case and the reduction is $25\%$ for $k = 20$. Thus, besides providing the guarantee on the optimal solution before even applying the Lloyd descent, the weighted-$D^2$ also leads to significant energy savings during the descent. If the sensor motion dynamics is the same in both cases, this also implies faster convergence to the optimal solution.

\begin{table*}[htbp]
\begin{center}
{\small
\hfill{}
    \begin{tabular}{|c|c|c|c|c|c|c|}
    \hline
\multicolumn{2}{|c|}{Case} & \multicolumn{2}{|c|}{Weighted-$D^2$ initial configuration} & \multicolumn{2}{|c|}{Uniform random initial configuration} & \% Improvement \\\hline
Sensors & Grid size & Initial coverage & Final coverage & Initial coverage & Final coverage & Initial coverage \\
($k$) & ($\varepsilon$) & ($H_{D^2})$ & ($H_{D^2}^\ast)$ & ($H_{U}$) & ($H_{U}^\ast$)  & ($\frac{H_{U} - H_{D^2}}{H_{U}}$)\% \\\hline
10 & 0.1 & 0.0235 $\pm$ 0.0023 & 0.0154 $\pm$ 0.0001 & 0.0372 $\pm$ 0.0085 & 0.0155 $\pm$ 0.0002  & 36.7 \\\hline
10 & 0.05 & 0.0236 $\pm$ 0.0024 & 0.0154 $\pm$ 0.0001 & 0.0353 $\pm$ 0.0115 & 0.0154 $\pm$ 0.0001  & 33.1 \\\hline
20 & 0.05 & 0.0121 $\pm$ 0.0011 & 0.0077 $\pm$ 0.0000 & 0.0179 $\pm$ 0.0054 & 0.0077 $\pm$ 0.0000  & 32.4 \\\hline
    \end{tabular}}
      \hfill{}
  \caption{Comparison of the means and the standard deviations of the initial coverage and the final coverage after the Lloyd descent across 50 runs for two types of initial sensor configurations for three scenarios. The last column shows the \% improvement in the average initial coverage value for the weighted-$D^2$ sampling over the uniform random sampling case.}
  \label{table:results1}
\end{center}
\end{table*}

\begin{table*}[htbp]
\begin{center}
{\small
\hfill{}
    \begin{tabular}{|c|c|c|c|c|}
    \hline
\multicolumn{2}{|c|}{Case} & \multicolumn{2}{|c|}{Average distance traveled per sensor during the Lloyd descent } & \% Improvement \\\hline
Sensors & Grid size & Weighted-$D^2$ initial configuration & Uniform random initial configuration & Average distance \\
($k$) & ($\varepsilon$) & ($d_{D^2})$ & ($s_{U}$) & ($\frac{d_{U} - d_{D^2}}{d_{U}}$)\% \\\hline
10 & 0.1 & 0.2281 $\pm$ 0.0512 & 0.3441 $\pm$ 0.0747  & 33.7 \\\hline
10 & 0.05 & 0.2159 $\pm$ 0.0485 & 0.3173 $\pm$ 0.0930  & 32.0 \\\hline
20 & 0.05 & 0.1633 $\pm$ 0.0287 & 0.2192 $\pm$ 0.0429  & 25.5 \\\hline
\end{tabular}}
      \hfill{}
  \caption{Comparison of the means and the standard deviations of the average distance traveled by the sensors during the Lloyd descent across 50 runs for two types of initial sensor configurations for three scenarios. The last column shows the \% improvement in the average distance traveled per sensor for the weighted-$D^2$ sampling over the uniform random sampling case.}
  \label{table:results2}
\end{center}
\end{table*}

\section{Conclusions and Future work}
\label{sec: conclusion}
In this paper we revisited the mobile sensor coverage problem formulated as a continuous locational optimization problem. The Lloyd descent algorithm yields a locally optimal solution with the guaranteed convergence. Researchers have extensively studied variations of this coverage problem and have proposed distributed versions of the variations of the Lloyd descent algorithm to achieve locally optimal solutions with guaranteed convergence. The quality of the final solution depends on the initial sensor configuration which is typically chosen uniformly at random. In this paper, we focus on the original coverage problem formulation and propose the weighted-$D^2$ sampling to choose the initial sensor configuration and show that it yields $O(\log k)$-competitive sensor coverage before even applying the Lloyd descent. We proved our result in two steps. In the first step, we established a close relationship between the sensor coverage problem and a suitably selected weighted $k$-means problem. In the second step, we extended the original result in \cite{Arthur:2007:KAC:1283383.1283494} to the weighted-$D^2$ sampling for the weighted $k$-means problem. Through extensive numerical simulations, we show that the initial coverage with the weighted-$D^2$ sampling is significantly lower than the uniform random initial sensor configuration. We also showed that the average distance traveled by the sensors to reach the final configuration through the Lloyd descent is also significantly lower with the weighted-$D^2$ sampling than that with the uniform random initial deployment. This also means considerable savings in the energy spent by the sensors during motion and faster convergence.

In future, we plan to extend this work to address guaranteed coverage for the variations of the coverage problem such as generic monotonic sensing performance functions. We also plan to extend the sampling procedure to include the learning of the density function.

\bibliographystyle{plain}

\bibliography{ref}

\section{Appendix: Proof of Theorem \ref{thm:k-means guarantee}}
\label{sec:appendix proof weighted k-means}

Let $P^o = \{p_1^o, p_2^o, \cdots, p_k^o\}$ denote an optimal solution to $\Phi(P)$. Thus, $\Phi_{\mathrm{OPT}} = \Phi(P^o)$.

\begin{lemma}
\label{lemma:single center expectation}
Let $p_a^o$ denote an arbitrary center from the optimal solution $P^o$ and let $A$ denote the cluster of $x_i$'s that are covered by $p_a^o$. Consider another clustering with just one center $x_i$ which is chosen from $A$ with the probability proportional to $w_i$. Then,
\begin{equation}\label{eqn:single center relation}
    \mathrm{E}[\phi(A)] = 2\phi_{\mathrm{OPT}}(A).
\end{equation}
\end{lemma}

\emph{Proof:} $p_a^o$ is the center of mass of $A$. Thus,
\begin{equation}
\label{eqn:center of mass}
    p_a^o = \frac{\sum_{i\in A} w_i x_i}{\sum_{i\in A} w_i}.
\end{equation}
\begin{eqnarray*}
  \mathrm{E}[\phi(A)] &=& \sum_{i\in A} \frac{w_i}{\sum_{j\in A} w_j} \sum_{j \in A} w_j \|x_i-x_j\|^2 \\
   &=& \sum_{i\in A} \frac{w_i}{\sum_{j\in A} w_j} \sum_{j \in A} w_j \|x_i - p_a^o - (x_j - p_a^o)\|^2 \\
   &=& \sum_{i\in A} \frac{w_i}{\sum_{j\in A} w_j} \sum_{j \in A} w_j \|x_i - p_a^o\|^2 + \\
   & & \sum_{i\in A} \frac{w_i}{\sum_{j\in A} w_j} \sum_{j \in A} w_j \|x_j - p_a^o\|^2 - \\
   & & \sum_{i\in A} \frac{w_i}{\sum_{j\in A} w_j} \sum_{j \in A} 2w_j (x_i - p_a^o)\cdot(x_j - p_a^o).
\end{eqnarray*}
In the last equation above, note that the first two terms on the right hand side are the same and each is reduced to $\sum_{i\in A} w_i\|x_i - p_a^o\|^2$. The third term is 0 by the definition of center of mass in Equation \ref{eqn:center of mass}. Therefore,
\begin{eqnarray*}
\mathrm{E}[\phi(A)] = 2\sum_{i\in A} w_i\|x_i - p_a^o\|^2 = 2\phi_{\mathrm{OPT}}(A).
\end{eqnarray*}
\qed

\begin{lemma}
Let $p_a^o$ denote an arbitrary center from the optimal solution $P^o$ and let $A$ denote the cluster of $x_i$'s that are covered by $p_a^o$. Consider an arbitrary clustering $\mathcal{C}$. If we add another center to $\mathcal{C}$ chosen from $A$ with weighted-$D^2$ sampling, then $\mathrm{E}[\phi(A)] \le 8\phi_{\mathrm{OPT}}(A)$.
\end{lemma}
\emph{Proof:}
Let $D(x)$ denote the shortest distance of $x$ to the already existing centers in $\mathcal{C}$. Given that we add a new center from $A$, the probability that we choose $x_i \in A$ is $\frac{w_iD(x_i)^2}{\sum_{i\in A}w_iD(x_i)^2}$. After choosing $x_i$, each $x_j \in A$ will contribute $w_j\min(D(x_j),\|x_i-x_j\|)^2$ to $\phi(A)$. Therefore,
\begin{equation}
\label{eqn:lemma arbitrary center 2}
    \mathrm{E}[\phi(A)] = \sum_{i\in A}\frac{w_iD(x_i)^2}{\sum_{i\in A}w_iD(x_i)^2}\sum_{j\in A}w_j\min(D(x_j),\|x_i-x_j\|)^2.
\end{equation}
By triangle inequality, $D(x_i) \le D(x_j) + \|x_i-x_j\|$ for all $x_i, x_j$. By squaring both sides, we have $D(x_i)^2 \le D(x_j)^2 + \|x_i-x_j\|^2 + 2\|x_i-x_j\|D(x_j)$. Note that $D(x_j)^2 + \|x_i-x_j\|^2 \ge 2\|x_i-x_j\|D(x_j)$. Therefore we have, $D(x_i)^2 \le 2D(x_j)^2 + 2\|x_i-x_j\|^2$. Multiplying by $\frac{w_j}{\sum_{j\in A}w_j}$ and summing over all $j \in A$, we get,
\begin{eqnarray*}
    D(x_i)^2 \le \frac{2}{\sum_{j\in A}w_j}\sum_{j \in A}w_j D(x_j)^2 \\
    + \frac{2}{\sum_{j\in A}w_j}\sum_{j \in A} w_j\|x_i-x_j\|^2
\end{eqnarray*}
Substituting in \ref{eqn:lemma arbitrary center 2},
\begin{eqnarray*}
    \mathrm{E}[\phi(A)] \le \sum_{i\in A}\frac{2w_i}{\sum_{j\in A}w_j}\frac{\sum_{j \in A}w_j D(x_j)^2}{\sum_{i\in A}w_iD(x_i)^2}\\ \times\sum_{j\in A}w_j\min(D(x_j),\|x_i-x_j\|)^2 \\
    + \sum_{i\in A}\frac{2w_i}{\sum_{j\in A}w_j}\frac{\sum_{j \in A} w_j\|x_i-x_j\|^2}{\sum_{i\in A}w_iD(x_i)^2}\\
    \times\sum_{j\in A}w_j\min(D(x_j),\|x_i-x_j\|)^2.
\end{eqnarray*}
In the first expression, we substitute $\min(D(x_j),\|x_i-x_j\|)^2 \le \|x_i-x_j\|^2$, and in the second expression, we substitute, $\min(D(x_j),\|x_i-x_j\|)^2 \le D(x_j)^2$.
\begin{eqnarray*}
  \mathrm{E}[\phi(A)] &\le& 4\sum_{i\in A}\frac{w_i}{\sum_{i\in A}w_i} \sum_{j\in A}w_j\|x_i-x_j\|^2\\
   &=& 8\phi_{\mathrm{OPT}}(A).
\end{eqnarray*}
The last step above follows from Lemma \ref{lemma:single center expectation}.
\qed

We have now shown that as long as our sampling chooses centers from each cluster from the optimal solution $P^o$, the cost metric is competitive. The lemma below bounds the total error.

\begin{lemma}
\label{lemma:induction argument}
Let $\mathcal{C}$ be an arbitrary clustering obtained by choosing some $x_i$'s as centers. Let $\mathcal{C}_{\mathrm{OPT}}$ be the clustering corresponding to the optimal solution $P^o$. A cluster from $\mathcal{C}_{\mathrm{OPT}}$ is ``uncovered'' if none of the $x_i$'s in that cluster are part of the centers of $\mathcal{C}$. Let $\mathcal{X}_u$ denote the set of ``uncovered'' points in these clusters. Also, let $\mathcal{X}_c = \mathcal{X} - \mathcal{X}_u$. Now suppose we add $t \le u$ random centers to $\mathcal{C}$ chosen with the weighted-$D^2$ sampling. Let $\mathcal{C}'$ denote the resulting clustering and let $\Phi'$ denote the corresponding cost metric. Then,
\begin{equation*}
E[\Phi'] \le \left( \Phi(\mathcal{X}_c) + 8\Phi_{\mathrm{OPT}}(\mathcal{X}_u) \right)\cdot(1 + H_t) + \frac{u-t}{u}\cdot\Phi(\mathcal{X}_u).
\end{equation*}
Here, $H_t$ denotes the harmonic sum, $1 + \frac{1}{2} + \cdots + \frac{1}{t}$.
\end{lemma}

The proof of the above lemma follows the exact argument in \cite{Arthur:2007:KAC:1283383.1283494} word to word and hence, it is not repeated here. Now we are ready to prove Theorem \ref{thm:k-means guarantee}.

\emph{Proof of Theorem \ref{thm:k-means guarantee}}: Let $\mathcal{C}$ denote the clustering after choosing the first center in Step (1) of the weighted-$D^2$ sampling. Let $A$ be the cluster from $\mathcal{C}_{\mathrm{OPT}}$ which contains this first center. After applying Lemma \ref{lemma:induction argument} with $t = u = k-1$ and $A$ being the only covered cluster, we obtain,
\begin{equation*}
E[\Phi] \le \left( \Phi(A) + 8\Phi_{\mathrm{OPT}} - 8\Phi_{\mathrm{OPT}}(A) \right)\cdot(1+H_{k-1}).
\end{equation*}
The result follows from Lemma \ref{lemma:single center expectation} and $H_{k-1} \le \ln k + 1$.
\end{document}